\documentclass[twoside]{article}
\usepackage{fleqn,espcrc2}
\usepackage{times}

\usepackage{graphicx}

\newcommand{\AmS}{{\protect\the\textfont2
  A\kern-.1667em\lower.5ex\hbox{M}\kern-.125emS}}
\title{The underdoped-overdoped transition in YBa$\bf_2$Cu$\bf_3$O$\bf_{x}$}

\author{J\"urgen R\"ohler \address{ 
        Universit\"at zu K\"oln, 
        Z\"ulpicher Str. 77, D-50937 K\"oln, Germany}%
        \thanks{E-Mail: abb12@uni-koeln.de (Juergen Roehler)}}

\begin{document}
    
\begin{abstract}
Oxygen doping in metallic YBa$_2$Cu$_3$O$_{x}$ induces 
quadrupolar ``$\alpha$-ortho'', and breathing ``$\beta$-ortho'' deformations of the 
CuO$_{2}$ planes. Breathing $\beta$-ortho deformations favour hybridizations of 
the $pd\sigma$ Cu$3d_{x^2-y^2}$--O$2p_{x,y}$ with the 
$pd\pi$ Cu$3d_{x,z},3d_{y,z}$--O$2p_{z}$ bands  
relaxing the confinement of the carriers in the overdoped regime, 
$x\geq 6.95$.
\vspace{1pc}
\end{abstract}


\maketitle

\section{INTRODUCTION}

The overdoped regime of YBa$\bf_2$Cu$\bf_3$O$\bf_{x}$ is accessible 
with heavy oxygen dopings close to $x$ = 7. The crossover from the 
underdoped to the overdoped regimes occurs around $x_{opt}$ = 6.92, 
nearly coinciding with a displacive structural phase 
transition \cite{KalLoe}. It is suggesting to 
associate the structural instability close to optimum doping  
with a kind of barrier limiting $T_{c}$. The 123 bi-layer cuprates 
exhibit comparatively low ``optimum'' $T_{c}$`s around 90 K, but 
chemical trends and pressure experiments 
imply that $T_{c}$  is not yet at its optimum value. 
In this communication we discuss  
a mechanism being possibly responsible for the barrier limiting 
$T_{c}$ of YBa$\bf_2$Cu$\bf_3$O$\bf_{x}$, and stabilizing the 
overdoped regime for $x\geq 6.95$.

\section{DOPING INDUCED LATTICE EFFECTS}

Oxygen doping of the parent cuprate YBa$\bf_2$Cu$\bf_3$O$\bf_{6}$ 
induces two displacive structural phase 
transitions: the first at the insulator metal transition around $x=6.42$  
transforming the tetragonal unit cell of the antiferromagnetic 
insulator into the orthorhombic symmetry of the metal, 
and  the second close to 
optimum doping, $x=6.95$, changing the orthorhombic 
unit cell from the $\alpha$-ortho to the $\beta$-ortho type. 

\subsection{$\alpha$-orthorhombic deformation}

In the underdoped regime ($6.42\leq x\approx 6.86$) doping is well 
established to increase the orthorhombicity, $2000(b-a)/(b+a)$. 
Positive axial strain, $\partial b/ \partial 
x \geq 0$ expands weakly the $b$-axis, while negative axial 
strain, $\partial a/ \partial x \leq 0$ 
compresses strongly the $a$-axis, cf. Fig.1 ({\it left}). 
The oppposite axial strains result 
in a {\it quadrupolar} instability in the orthorhombic basal plane, 
labelled $\alpha$-ortho.

\subsection{$\beta$-orthorhombic deformation}

In the overdoped regime ($6.95\leq x \approx 7$) the $b$-axis strain 
changes its sign, hence doping shrinks both, $a$ and $b$, resulting in a {\it 
breathing} instability of the orthorhombic basal plane, labelled 
$\beta$-ortho, cf. Fig.1 ({\it right}). Since both, $b$- and $a$-axis 
strains, are negative in all samples, independent on their various 
routes of chemical preparation, the planar breathing instability has to 
be considered as a generic property of the overdoped regime. 

\begin{figure}[thb]
\begin{center}\leavevmode
\includegraphics[width=1.\linewidth]{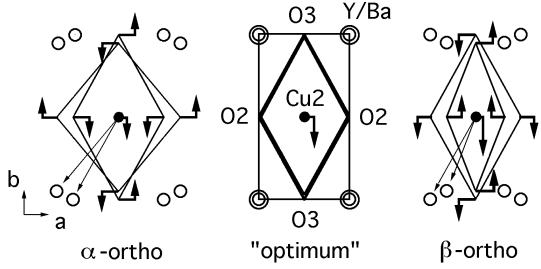}
\caption{Incremental deformations of the copper-oxygen layers upon doping in the 
underdoped ({\it left}), optimum doped ({\it middle}), and 
overdoped regimes ({\it right}), viewed from the top, see text. 
Note that the ``planes'' are stacked layers, 0.25-0.29 {\AA} thick.
Cu2 is close to 
Ba--O1(Apex), and O2,O3 are close to Y--Y. The {\it a}-axis O2 
are about 0.007 {\AA} closer to Cu2 than 
the {\it b}-axis O3. The out-of-plane displacements are indicated by thick arrows. 
``Up'' is towards the Y-layer, and ``down'' towards the Ba--O1 layer.
The long arrows indicate the orthorhombic shear along (110). 
} 
\label{fig1}
\end{center}
\end{figure}

\subsection{$c$-axis stress}
Doping compresses strongly the $c$-axis, 
usually described by a straight 
throughout all regimes up to $x=7$. 
Samples synthezised in the absence of carbonate however 
were shown to exhibit a significant minimum in $c(x)$ close to 
the onset of the overdoped regime \cite{KalLoe}. 
Raman spectroscopy and Y-EXAFS 
confirm the $c$-axis anomaly also in other samples: 
the in-phase O2,O3 mode (out-of-plane) softens abruptly  
in the overdoped regime, while the static 
O2,3--Cu2 interlayer spacing (``dimple'') 
increases discontinously by $\approx 0.015$ {\AA} \cite{Roe}. 

The strong contraction of $c$ is nearly completely 
attributable to the compression of the Cu2--O1(Apex) bonds 
reflecting the charge transfer from the chains to the planes. 
The Cu2 atoms may be seen to be pulled out the CuO$_{2}$ planes.
Notably thereby the average Y--O2,3 interlayer spacings remain unaffected.

Doping across the underdoped--overdoped phase boundary 
continues compressing the Cu2--O1(Apex) bonds. 
In the overdoped regime however the average O2,3-layers experience a significant 
{\it repulsion} from the Cu2-layers, strongly contrasting with 
the underdoped regime. In some samples the Y--O2,3 interlayer 
spacing remains nearly unaffected, 
and as a result $c(x)$ develops a minimum \cite{KalLoe}.

\section{CORRELATED DISPLACEMENTS IN THE CuO$_{2}$ PLANE}

We have shown from Y-EXAFS that the 
Y--Cu2 bond lengths are almost independent on doping \cite {Roe}. 
This suggests 
an ``umbrella mode'' to describe the static correlations 
between the Cu2 out-of-plane, 
and the O2, O3 in-plane displacements \cite{Roe99}. A refinement of 
the ``umbrella mode'' model may also explain 
the $\alpha$- and  $\beta$-ortho 
deformations in the under- and overdoped regimes, respectively. 
The lengths of the rigid semicovalent
in-plane copper-oxygen bonds can be safely assumed to be almost independent 
on doping, too. 
Then collapsing the ``umbrella'' moves the planar oxygens O2,3
out of their planes. As schematically shown in Fig.1 
({\it left}) the {\it quadrupolar} $\alpha$-ortho 
instability pushes the O3 atoms beyond,
and the O2 atoms beneath their planes. These shifts in opposite directions 
along $c$ tend to cancel (depending slightly on the orthorhombicity),  
and the plane dimpling 
is solely determined by the Cu2 out-of-plane displacement. 
On the other hand 
side (Fig.1, {\it right}) the {\it breathing} $\beta$-ortho instability 
pushes both, O2 and O3 atoms, beyond their planes, thus increasing 
the dimpling not only by the Cu2 out-of-plane shift, but also by 
the oppositely directed O2,O3 out-of-plane shifts.

\section{ELECTRONIC STRUCTURE EFFECTS}

Out-of-plane displacements of the planar copper and oxygens 
were shown to have important effects 
on the electronic band structure near $E_{F}$. In particular they  
determine the strengths and symmetries of the 
interlayer hoppings \cite{And}. Remote hybridization 
of the $\sigma$ Cu$4s$ orbitals with the $\sigma$ Cu$3d_{x^2-y^2}$--
O$2p_{x,y}$ band provides interlayer hoppings with $d$-symmetry, 
exhibiting maxima at ($\pi$,0) and  
(0,$\pi$) in $k$-space. 
The strength of the $sdp$ hybridization 
is mainly controlled by the length of Cu2--O1 (Apex) bond. 

Hybridizations of the the $pd\pi$ 
Cu$3d_{x,z}$--O$2p_{z}$, Cu$3d_{y,z}$--O$2p_{z}$ 
with the $pd\sigma$ Cu$3d_{x^2-y^2}$--O$2p_{x,y}$ bands  
become feasible through Cu2 out-of-plane positions. 
Once activated, $pd\pi$-$pd\sigma$ hybridizations 
repell the O2,3 from the Cu2 plane \cite{And}.

We conclude, that the {\it quadrupolar} $\alpha$-ortho instability 
ought to suppress hybridizations of the $pd\pi$ with the $pd\sigma$
bands. The {\it breathing} 
$\beta$-ortho instability however ought to favour them 
and to provide a possible mechanism relaxing the confinement of 
the carriers, thus pushing the system into the overdoped regime.

\bigskip
I acknowledge the support of the ESRF through the projects HE731 and 
HE516.

\end{document}